\documentclass[aps,prb, twocolumn, showpacs,superscriptaddress]{revtex4-1}

\usepackage{amssymb,amsfonts,amsmath} 
\usepackage{graphicx,epsfig,psfrag}
\usepackage{color}
\usepackage{natbib}
\usepackage{url}
\usepackage[breaklinks=true]{hyperref}
\usepackage{mathtools}
\usepackage{subfigure}
\hypersetup{
        colorlinks=true,
        citecolor    = blue
}
\usepackage{bookmark}
\usepackage{epic}
\usepackage{longtable}

\usepackage{bbm}
\usepackage{ulem}
\usepackage{braket}
\usepackage{xcolor}

\begin{document} 

\title{Influence of phonon-assisted tunneling on the linear thermoelectric transport through molecular quantum dots}

\author{A.\ Khedri}
\affiliation{Institut f{\"u}r Theorie der Statistischen Physik, RWTH Aachen University 
and JARA---Fundamentals of Future Information
Technology, 52056 Aachen, Germany}
\affiliation{Peter Gr\"unberg Institut and Institute for Advanced Simulation, Research Centre J\"ulich, 
52425 J\"ulich, Germany} 
\author{V.\ Meden} 
\affiliation{Institut f{\"u}r Theorie der Statistischen Physik, RWTH Aachen University 
and JARA---Fundamentals of Future Information
Technology, 52056 Aachen, Germany}
\author{T.A.\ Costi}
\affiliation{Peter Gr\"unberg Institut and Institute for Advanced Simulation, Research Centre J\"ulich, 
52425 J\"ulich, Germany}

\begin{abstract} 
We investigate the effect of vibrational degrees of freedom on the linear thermoelectric transport through a single-level quantum dot described by the spinless Anderson-Holstein impurity model.
To study the effects of strong electron-phonon coupling, we use the nonperturbative numerical renormalization group approach.  We also compare our results, at weak to intermediate coupling,
with those obtained by employing the functional renormalization group method, finding good agreement in this parameter regime.   
When applying a gate voltage at finite temperatures, the inelastic scattering processes, induced by phonon-assisted tunneling, result in an interesting interplay between electrical and thermal transport.  
We explore different parameter regimes and identify situations for which the thermoelectric power as well as the dimensionless 
figure of merit are significantly enhanced via a Mahan-Sofo type of mechanism. We show, in particular, that this occurs at strong electron-phonon coupling and in the antiadiabatic regime.
\end{abstract}

\date{\today} 
\maketitle

\section{Introduction}
\label{sec:intro}

Molecular quantum dots can be considered as potential candidates for the interconversion of heat and electrical energy,
with possible applications to cooling or energy harvesting at the nanoscale.\cite{Galperin07} 
From this perspective, studying the thermoelectric transport through such nanostructures is of crucial importance as it can 
help us to identify scenarios for which the thermoelectric efficiency can be enhanced. 
However, there are both experimental\cite{Reddy17} and theoretical\cite{Segal16} challenges in understanding the thermoelectric properties of 
such systems. On the experimental side there is the technical challenge of applying a small temperature gradient across a nanoscale device and measuring the resulting thermovoltage,
\cite{Reddy17}, while on the theoretical side a major challenge is the inclusion of correlation effects in the calculation of thermoelectric 
transport. Interactions such as the on-site Coulomb repulsion or the local electron-phonon interaction are known to be  important for nanoscale systems. 
For example, the Kondo effect \cite{Hewson97}  can lead to a drastic modification of the low-temperature conductance of molecular junctions. \cite{Park02} 

As correlation effects have prominent consequences on the transport properties of molecular quantum dots, we need sophisticated 
many-body methods in order to address them in a satisfactory manner.
The numerical renormalization group (NRG) provides an accurate nonperturbative description of these properties in all parameter regimes and over the whole temperature range.\cite{Wilson75,Krishnamurthy80,GonzalezBuxton98,Bulla08} It can deal with arbitrarily complicated local interactions, including local Coulomb and electron-phonon interactions. 
Within the framework of the single impurity Anderson model, NRG has been applied to study the thermoelectric transport properties of strongly interacting quantum dots for 
both repulsive\cite{Costi10} as well as attractive Coulomb interactions\cite{Andergassen11} on the dot. Attractive interactions, in particular, were found to provide a
mechanism for enhanced thermoelectric power and efficiency in molecular quantum dots.\cite{Andergassen11}
The effect of a short-range Coulomb interaction at the contact points between the dot and the leads (interacting resonant level model) 
has also been studied within the approximate functional renormalization group method (FRG), both in and out of equilibrium  (steady state).\cite{Kennes13} 

In the quest to find general criteria for the best thermoelectric material, Mahan and Sofo \cite{Mahan95} realized that a narrow distribution of the spectral weight of the quasiparticles involved in the transport can result in a substantial thermoelectric efficiency.
At first glance, repulsive interactions on a quantum dot, resulting in a sharp Kondo resonance, seem to be a realization of such a situation.
However, the spin Kondo resonance is generally well pinned fairly close to the Fermi level, and 
inducing some asymmetry in the spectral function to enhance the low-temperature Kondo-induced thermopower either by applying a gate voltage or an external magnetic field turn out to have small effects,\cite{Costi10,Merker16} at least within the simplest model, the spin degenerate Anderson impurity model.\footnote{Orbitally degenerate Anderson models, used to describe heavy fermion materials, can result in a larger asymmetry of the Kondo resonance about the Fermi level and consequently can have a larger Kondo induced thermopower at low temperatures, see Refs.~\protect\onlinecite{Zlatic05,Paschen13}.}
The charge Kondo physics realized in quantum dots with  attractive interaction, on the other hand, can result in significant enhancement of the Seebeck coefficient through a large polarization of the spectral function caused by a small charge splitting (gate voltage).\cite{Andergassen11}  

In this paper, we explore a different route to enhanced thermoelectric efficiency, by considering the effects not of local
Coulomb correlations, but of local vibrational modes of the molecular device.
As vibrations are inevitable features of real molecular quantum dots,\cite{Park02} we want to identify the signatures of the vibrational modes on the linear transport properties through such devices and also characterize the regime of parameters for which vibrational effects lead to enhanced thermoelectric efficiency. 
For this purpose we take the spinless Anderson-Holstein model as a simple model of a molecular junction, and   
investigate its linear thermoelectric properties at finite temperatures within the NRG. Since the latter approach is nonperturbative in all interactions,\cite{Bulla08}
it includes all tunneling processes between the dot and the leads and can therefore be used to access both the low-temperature strong coupling regime at $T\ll \Gamma_{\rm eff}$,
and the high-temperature perturbative regime $T\gg\Gamma> \Gamma_{\rm eff}$, where $\Gamma_{\rm eff}<\Gamma$ is the renormalized tunneling rate between the molecular quantum dot and the leads and $\Gamma$ is the bare one (see Sec.~\ref{sec:method} for the precise definitions).
In addition, at weak to intermediate electron-phonon coupling, we compare the transport coefficients calculated within NRG with those obtained from FRG.

Many studies have focused attention on the spinful version of the Anderson-Holstein model,\cite{Hewson02,Jeon03}  
which includes a local Coulomb repulsion on the dot, and on the resulting competition between Kondo physics and electron-phonon effects.
For this model, and variants thereof,\cite{Galperin07,Koch04,Mengxing13} a large number of results have been obtained, including  
the linear \cite{Cornaglia04,*Cornaglia05}  and  nonlinear \cite{Paaske05,Hyldegaard94,Koenig96,Han06,*Han10,Schiro09,Mitra04,Laakso14} electrical conductance,  
the thermopower in the perturbative high-temperature limit $T\gg\Gamma$,\cite{Koch04} and other thermoelectric properties.\cite{Lijnse10,Perroni14,Zimbovskaya16,Ren12}  
In contrast, previous studies of the spinless Anderson-Holstein model have mainly focused on renormalization effects on the low-energy scale \cite{Sherrington75,Hewson79,*Hewson80,Eidelstein13,Khedri17} 
and on the electrical conductance. \cite{Muehlbacher08,Koch11,Huetzen12,Jovchev13} To the best of our knowledge,  
the effects of electron-phonon coupling on the other transport coefficients (thermopower, thermal conductance) and on the dimensionless figure of merit as well as the Lorenz number  
have not been previously addressed. The main aim of the present paper is to fill this gap and to elucidate in detail the signatures of phonon-assisted tunneling in thermoelectric properties,
without the added complication of Kondo physics in the spinful version of this model.
 
The outline of the paper is as follows: In Sec.~\ref{sec:method} we introduce the model, outline very briefly the NRG and FRG methods,
and describe how finite-temperature transport is calculated within these approaches.
A more extensive description of the methods themselves, in the context of the present model is given in Ref.~\onlinecite{Khedri17}. While in the latter paper, following the pioneering study of Ref.~\onlinecite{Eidelstein13},
we elucidated in detail the evolution of the low-energy scale of the model from the adiabatic to the antiadiabatic regime and from weak to intermediate electron-phonon couplings,
using NRG, FRG, and perturbation theory, and compared also the $T=0$ spectral functions within these methods, in the present paper we focus our attention on finite temperature thermoelectric transport properties for molecular quantum dots strongly coupled to leads. Our results for these, at temperatures above and below the relevant low-energy scale $\Gamma_{\rm eff}$, are presented in Sec.~\ref{sec:results}, and we conclude with an outlook in Sec.~\ref{sec:summary}. 
In the Appendixes, we describe the details of the FRG calculations for finite temperature thermoelectric transport (Appendix~\ref{subsec:frg-finite-T}), indicate the 
convergence tests used for the NRG calculations (Appendix~\ref{subsec:NRG-parameters}), present 
some additional results for the dependence of the dimensionless figure of merit on the phonon frequency (Appendix~\ref{subsec:additional-results}), and
show results for the coupling strength, temperature, phonon frequency, and gate voltage dependence of the  molecular dot spectral function (Appendix~\ref{subsec:spectral_function}). 
For completeness, and to illustrate the applicability of the NRG approach also in the high-temperature perturbative limit at $T\gg\Gamma>\Gamma_{\rm eff}$, we discuss in Appendix~\ref{subsec:weakGamma}
the evolution of the thermopower (versus gate voltage) from its high-temperature perturbative limit to its low-temperature strong coupling 
behavior at $T\lesssim \Gamma_{\rm eff}$.

\section{Model, methods and transport calculations}
\label{sec:method}
                
We focus on the simplest possible model to capture the vibrational effects of a molecule in a tunnel junction, the so-called spinless Anderson-Holstein model,
\begin{align}
H = &\sum_{\alpha=1}^{2} \sum_k \varepsilon_k c_{\alpha,k}^\dag c_{\alpha,k} 
+\frac{t}{\sqrt{N_{\rm sites}}} \sum_{\alpha=1}^{2} \sum_k \left( d^\dag c_{\alpha,k} + \mbox{H.c.}  \right) \nonumber\\
&+\epsilon_0 d^\dag d + \omega_0 b^\dag b + \lambda d^\dag d (b^\dag +b),
\label{eq:modelhamiltonian}
\end{align}
where $\epsilon_0$ is the energy of the molecular level, $\omega_0$ is the local phonon frequency, $\lambda$ is the strength of the electron-phonon coupling, and $t$ is the tunneling amplitude to the two leads, each of which is represented by a one-dimensional tight-binding chain with  $N_{\rm sites}$ lattice sites.
Due to polarization effects induced by the electron-phonon interaction, 
the particle-hole symmetric point of the Hamiltonian is shifted from the Fermi level $\epsilon_0=\epsilon_F=0$ to  $\epsilon_0=E_{P}$, 
where $E_p=\lambda^2/\omega_0$ is the polaronic shift.\footnote{Under the particle-hole transformation 
$d\to d^{\dagger}$, $b\to -b -\lambda/\omega_0$ with particle-hole symmetric leads we have that $H(\epsilon_0)\to H'=H(2E_{\rm p}-\epsilon_0)+(\epsilon_0-E_{\rm p})$, so for $\epsilon_0=E_{\rm p}$, $H=H'$.}
$E_{\rm p}/\omega_0$ may also be interpreted as the average number of phonons involved in the formation of a local polaron.\cite{Hewson80,Ingersent15} 
The quantity $\tilde{\varepsilon}_0=\epsilon_0-E_p$ is a measure of the deviation of the local level position from its particle-hole symmetric value and can be regarded as a gate voltage $-eV_g=\tilde{\varepsilon}_0$.
We shall henceforth parametrize all results by $\tilde{\varepsilon}_0$ rather than the bare level position $\epsilon_0$.
We shall consider the reservoirs (leads) to be structureless with a constant density of states $\rho_0(\omega)=1/(2D)$ with $D=1$ the half bandwidth.
The molecular level in (\ref{eq:modelhamiltonian}) couples to the left and right reservoirs with equal strength, resulting in a bare total level width of $\Gamma=2\pi\rho_0t^2$.
We use $\Gamma=10^{-4}D$ in all calculations. The low-temperature behavior of this model has been studied in depth in Refs.~\onlinecite{Hewson79,Hewson80,Eidelstein13} 
and also in our previous work where we used NRG and FRG \cite{Khedri17} and which we here extend to finite temperature transport. Due to its simplicity, the spinless Anderson-Holstein 
model is only expected to qualitatively capture some aspects of a real molecular quantum dot at low temperatures. At higher
temperatures (e.g., at $T\gg \max\{\Gamma,\omega_0\}$) additional complexities, not contained in the above simple model, such as additional molecular levels or anharmonic effects,
may become important and invalidate even a qualitative description in terms of the spinless Anderson-Holstein model. 
Hence, we will mainly focus on low temperatures, where also the most interesting many-body effects manifest themselves.

We are generally interested in the flow of charge and heat  through a vibrating molecule coupled to reservoirs held at different temperatures and chemical potentials.
In the linear response regime, all the transport coefficients of the model (\ref{eq:modelhamiltonian}) can be expressed in terms of the moments of the molecular dot spectral function $A(\nu)$ \cite{Jauho94},
\begin{align}
I_{n}(T)=-\pi\Gamma&\int_{-\infty}^{\infty}d\nu\nu^nA(\nu)(\partial_{\nu}f)_{T} ,
\label{eq:transportintegrals}
\end{align}
where $n=0,1,2$ and $f(\nu)$ is the Fermi function at temperature $T$. In particular,
 the electrical conductance $G(T)$, thermoelectric power (Seebeck coefficient) $S(T)$, and the electronic contribution to the thermal conductance $\kappa_e(T)$ can be calculated via
\begin{align}
G(T)=\frac{e^2}{h}I_{0}(T),\label{eq:GT}\\
S(T)=-\frac{1}{e}\frac{I_1(T)}{TI_0(T)}\label{eq:ST},\\
\kappa_e(T)=\frac{1}{hT}\bigg[I_2(T)-\frac{I_1^2(T)}{I_0(T)}\bigg],
\label{eq:KeT}
\end{align}
with $e$ and $h$ denoting the electric charge and Planck's constant, respectively.

Within the NRG approach to Eq.~(\ref{eq:modelhamiltonian}), described in more detail in Ref.~\onlinecite{Khedri17}, one  
obtains the eigenstates and eigenvalues of $H$ on all energy scales by an iterative diagonalization procedure involving a set of finite-size (or truncated)  
Hamiltonians $\mathcal{H}_M$, $M=0,1,\dots$. From these, one can then construct all equilibrium thermodynamic, dynamic, and linear transport quantities.\cite{Wilson75,Krishnamurthy80,GonzalezBuxton98,Bulla08}
Specifically,  we calculate the $n$-th moment of the spectral function (\ref{eq:transportintegrals}) at finite temperatures
following the best shell approach described in 
Refs.~\onlinecite{CampoOliveira05} and \onlinecite{Oliveira09}, namely, at temperature $T$, we find the corresponding 
best shell $M$ and use the information from this shell to evaluate
\begin{align}
I_{n}(T)=\frac{\pi\Gamma}{\mathcal{Z}_{M}(T)}\sum_{l,l^\prime=1}^{N_s}
\frac{|\bra{l^{\prime}}d^{\dag}\ket{l}|^2}{e^{-\beta E^{M}_l}+e^{-\beta E^{M}_{l^\prime}}}(E^M_l-E^M_{l^\prime})^n .
\label{eq:NRGmoments}
\end{align}
In the above, $\mathcal{Z}_M$ is the partition function associated with the truncated Hamiltonian $\mathcal{H}_M$ for a Wilson chain of length $M$ at an inverse temperature $\beta=1/T$ with $\{E^{M}_l\}$
the set of the $N_s$ lowest-lying  eigenvalues of  $\mathcal{H}_M$ and $\{\ket{l}\}$ the corresponding set of eigenvectors. In practice, we use a logarithmic discretization parameter of $\Lambda=4$
and average the results over $N_z=4$ realizations of the bath.\cite{Oliveira94,CampoOliveira05}

The Matsubara FRG formalism discussed in our previous work can be extended to finite temperatures following the procedure presented in Ref.~\onlinecite{Enss05}.
Within first-order truncated FRG, we calculate the self-energy $\Sigma(i\nu_n)$ at fermionic Matsubara frequency $\nu_n$
up to linear order in the effective-phonon-mediated electron-electron interaction ($\propto \lambda^2$).
However, due to the RG resummation, the results go well beyond the lowest order perturbation theory and also they preserve the particle-hole symmetric properties, in contrast to plain perturbation theory. 
The technical details of the method are discussed in Appendix~\ref{subsec:frg-finite-T}.
Knowing the molecular dot propagator
\begin{align}
G_{\rm mol}(i\nu_n)=[i\nu_n-\epsilon_0+i\Gamma\operatorname{sgn}(\nu_n)-\Sigma(i\nu_n)]^{-1}, 
\label{eq:molpropagator}
\end{align}
we use the continued fraction representation of the Fermi function to calculate the $n$-th moment of the spectral function without the analytic continuation to the real axis
\begin{align}
I_n=&(i)^{n-1}\frac{\pi\Gamma}{\beta}\sum_{p=1}^{M_p}\sum_{s=\pm}R_{p}\partial_\nu\big[\nu^n G(i\nu)\big]\Big|_{\nu=s\frac{z_{p}}{\beta}}\nonumber\\
&+\Gamma\delta_{n,2}\bigg[\Gamma-\frac{2\pi}{\beta}\sum_{p=1}^{M_p}{R_{p}}\bigg],
\label{eq:CFmoments}
\end{align}
with $M_p$ poles at positions $\pm i z_{p}/\beta$ and residues $R_{p}$ calculated as proposed in Refs.~\onlinecite{Ozaki07} and \onlinecite{Karrasch10}.
At low temperatures, we can also calculate the moments of the spectral function using the Sommerfeld expansion (see Appendix~\ref{subsec:frg-finite-T}).\cite{Kennes14}
\section{Results}
\label{sec:results}
In Sec.~\ref{subsec:transport properties} we present results for the temperature dependence of the various transport coefficients Eqs.~(\ref{eq:GT})-(\ref{eq:KeT}) 
of the spinless Anderson-Holstein model for different parameters ($\lambda/\omega_0$, $\tilde{\varepsilon}_{0}/\Gamma$, and $\omega_0/\Gamma$), while
in Sec.~\ref{subsec:fom} we likewise present results for the temperature and parameter dependence of the power factor, the Lorenz number, and the figure of merit.
As all the mentioned thermoelectric quantities are related to the spectral function [see Eqs.~(\ref{eq:GT})-(\ref{eq:KeT})], 
we trace back some of the trends to the behavior of the molecular spectral function presented in Appendix~\ref{subsec:spectral_function}. 

Throughout this section, we show the aforementioned quantities as a function of the reduced temperature $T/\Gamma_{\rm eff}$, where $\Gamma_{\rm eff} <\Gamma$ 
is the renormalized low-energy scale describing the rate at which tunneling processes occur between the dot and the leads at zero gate voltage
and zero temperature. \cite{Khedri17}
As in Ref.~\onlinecite{Khedri17}, we define $\Gamma_{\rm eff}$ in terms of the local $T=0$ charge susceptibility 
at zero gate voltage via $\Gamma_{\rm eff}=1/\pi \chi_c$, where $\chi_c=-\frac{d n_{d}(\tilde{\varepsilon}_0)}{d\tilde{\varepsilon}_0}|_{\tilde{\varepsilon}_0=0}$, and $n_d(\tilde{\varepsilon}_0)$
is the occupancy of the molecular level. This emergent low-energy scale $\Gamma_{\rm eff}$, the phonon frequency $\omega_0$, and the gate voltage $\tilde{\varepsilon}_0$ are the competing scales and they play a crucial role to understand the thermoelectric transport. 

In the following, we shall also mainly be interested in the antiadiabatic regime $\omega_0 \gg \Gamma$ 
where renormalization effects are most pronounced. In this case we have $\Gamma_{\rm eff} \ll \Gamma \ll \omega_0$ and we expect interesting temperature dependences in several temperature ranges 
defined by the disparate energy scales $\Gamma_{\rm eff}, \omega_0$ and the gate voltage  $\tilde{\varepsilon}_0$.  In the adiabatic regime  $\omega_0\ll \Gamma$, the physics is that of the noninteracting model 
and the only relevant temperature scale is $\Gamma$.  In the antiadiabatic limit, we discuss the comparison of NRG results with the corresponding FRG ones for a given 
intermediate coupling strength in Sec.~\ref{subsubsec:frg-comp-antiadiabatic}. The values of $\Gamma_{\rm eff}/\Gamma$ for the couplings used below are listed in Table~\ref{tab:gamma_eff} for the antiadiabatic case of $\omega_0=5\Gamma$. 

\begin{table}
\begin{ruledtabular}
\begin{tabular}{cccccc}
 $\lambda/\omega_0$ & 0.2 & 0.5 & 1.0 & 2.0 & 3.0 \\
 $\Gamma_{\text{eff}}/\Gamma$ & 0.975 & 0.851 & 0.503 & 0.039 & 0.00025\\
\end{tabular}
\end{ruledtabular}
\caption{\label{tab:gamma_eff} Dependence of $\Gamma_{\rm eff}/\Gamma$ on $\lambda/\omega_0$
 for $\omega_0=5\Gamma$.}
\end{table}
\subsection{Electrical conductance, thermopower and thermal conductance}
\label{subsec:transport properties}
\begin{figure*}[t]
 \centering
 \includegraphics[width=1.0\linewidth]{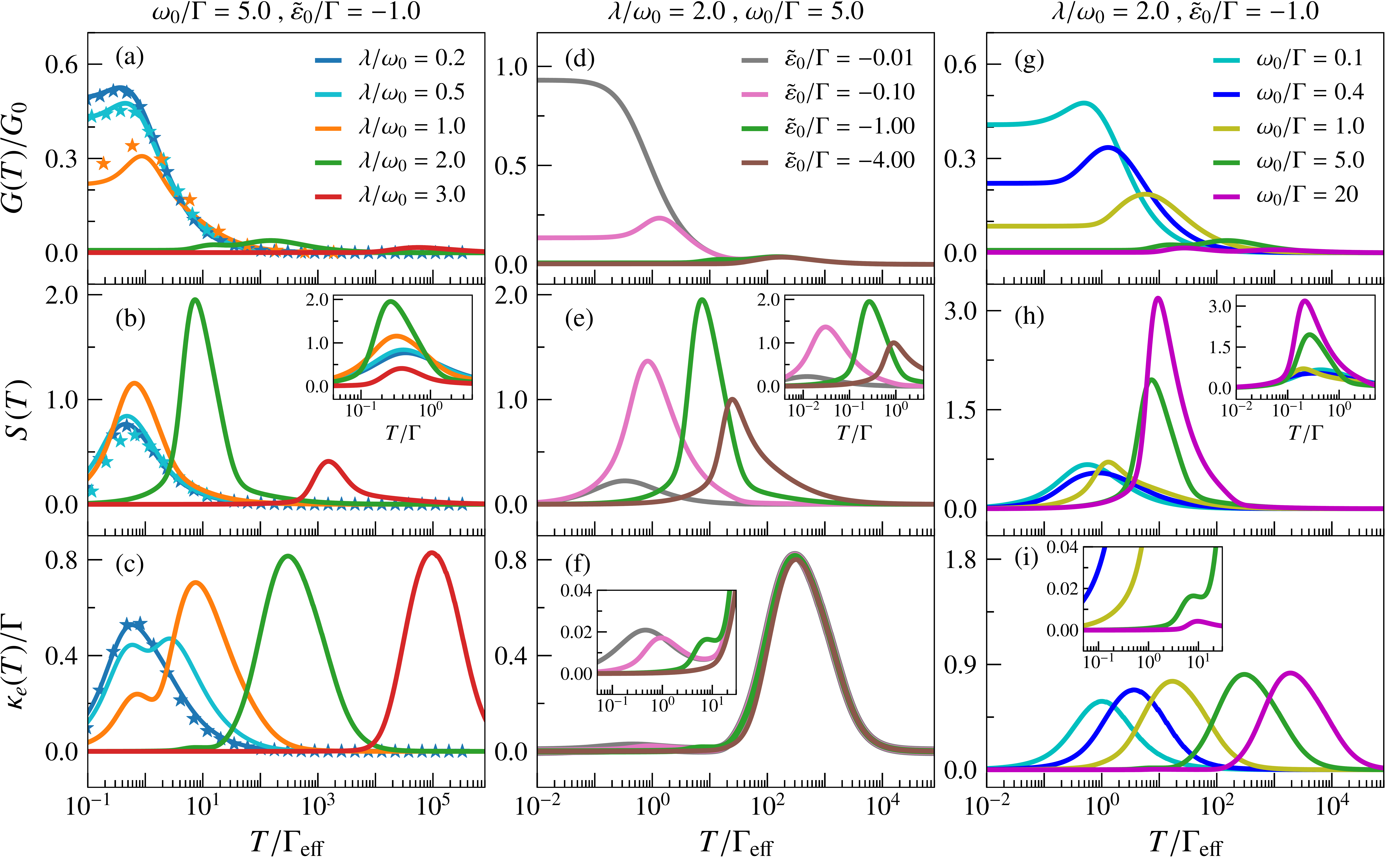}
 \caption{(Solid lines) NRG results for the temperature dependence of the normalized electric conductance $G/G_0$ ($G_0=e^2/h$), the thermopower $S$ (in units of $k_{\rm B}/e$), and the normalized electronic contribution to the thermal conductance $\kappa_e/\Gamma$ (in units of $k_{\rm B}/h$). (a)-(c) Evolution with increasing electron-phonon coupling  at a given phonon frequency $\omega_0/\Gamma=5$
 and level position $\tilde{\varepsilon}_0=-\Gamma$. Selected FRG results (stars) [using Eq.~(\ref{eq:CFmoments})] 
  at weak to intermediate couplings serve as checks on the NRG results. (d)-(f) Dependence on level position (gate voltage) for a given coupling 
 $\lambda/\omega_0=2.0$ and a fixed frequency $\omega_0/\Gamma=5$. (g)-(i) Evolution from the adiabatic to the antiadiabatic limit for $\lambda/\omega_0=2.0$ and $\tilde{\varepsilon}_0=-\Gamma$.}
 \label{fig:GSK}
\end{figure*}
In this section, we discuss the temperature dependence of the transport coefficients as a function of the electron-phonon coupling $\lambda$ (at fixed gate voltage $\tilde{\varepsilon}_0$ and phonon frequency $\omega_0$),
of the gate voltage $\tilde{\varepsilon}_0$ (at fixed coupling $\lambda$ and frequency $\omega_0$), and of the phonon frequency $\omega_0$ (at fixed coupling $\lambda$ and gate voltage $\tilde{\varepsilon}_0$). Results for these
three cases are shown in Figs.~\ref{fig:GSK}(a)-\ref{fig:GSK}(c), Figs.~\ref{fig:GSK}(d)-\ref{fig:GSK}(f) and Figs.~\ref{fig:GSK}(g)-\ref{fig:GSK}(i), respectively, and will be discussed in Secs.~\ref{subsubsec:GSK-coupling-dep}-\ref{subsubsec:GSK-frequency-dep}.

\subsubsection{Varying the electron-phonon coupling}
\label{subsubsec:GSK-coupling-dep}
In Figs.~\ref{fig:GSK}(a)-\ref{fig:GSK}(c), we show the temperature dependence of the transport coefficients
for different electron-phonon couplings at a fixed gate voltage, $\tilde{\varepsilon}_0=-\Gamma$, and a fixed phonon frequency in the antiadiabatic regime, $\omega_0=5\Gamma$. \footnote{The choice of these parameters is motivated by the desire, on the one hand, to be away from the particle-hole symmetric point $\tilde{\varepsilon}_0=0$, since the thermopower vanishes exactly there, and, on the other hand, to be in the antiadiabatic regime where, as explained above, the most interesting temperature dependences are expected.}

\paragraph{Electrical conductance}
While the electrical conductance $G$ at $T=0$ and particle-hole symmetry ($\tilde{\varepsilon}_0=0$) is pinned to its unitary value $G_0=e^2/h$ for all coupling strengths,\cite{Eidelstein13} at finite gate voltage,  as shown in Fig.~\ref{fig:GSK}(a),
it is strongly suppressed with increasing electron-phonon coupling. This results from the suppression of the spectral weight at the Fermi level with increasing coupling for finite gate voltages [see Fig.~\ref{fig:spectral_function}(a) in Appendix~\ref{subsec:spectral_function}]. 
At finite temperatures, the electrical conductance shows the typical behavior for resonant tunneling 
at finite gate voltages, an activated behavior at low temperatures with a maximum at a temperature related to the gate voltage and a decrease beyond this temperature.  
At still higher temperatures, as we approach the strong-coupling regime ($\lambda/\omega_0>1$), 
the electrical conductance develops another (small) maximum at a temperature related to $\omega_0$ showing that electrons can also tunnel by creating phonon excitations. 
The same observation holds for the thermal conductance, to be discussed below, which also shows a peak at a temperature $T$ related  to $\omega_0$ [Fig.~\ref{fig:GSK}(c)].
This feature arises from the multiphonon satellite peaks at $\nu\approx\tilde{\varepsilon}_0\pm n\omega_0, n=1,2,\dots$ in the spectral function at strong coupling 
[see Figs.~\ref{fig:spectral_function}(a) and \ref{fig:spectral_function}(b) in Appendix~\ref{subsec:spectral_function} and Ref.~\onlinecite{Jovchev13}].

\paragraph{Seebeck coefficient}
The Seebeck coefficient $S$, which probes the asymmetry of the spectral function about the Fermi level within the Fermi window
$|\omega|\lesssim T$, first increases with the strength of the electron-phonon coupling and then  decreases [see Fig.~\ref{fig:GSK}(b)] achieving a maximum at $\lambda/\omega_0\approx 2$ for the chosen $\omega_0/\Gamma=5$.
We may qualitatively understand these trends from the dependence of the spectral function on the coupling strength in Fig.~\ref{fig:spectral_function}(a) of Appendix~\ref{subsec:spectral_function} as follows; at small values of the coupling, 
most weight in the spectral function is carried by the central peak (located at $\nu\approx\tilde{\varepsilon}_0$) close to the Fermi level, resulting in a small asymmetry in the spectral function and a correspondingly small
thermopower. Increasing the coupling to values of order $\lambda/\omega_0\approx 1$ results in a gradual transfer of spectral weight from the central peak to 
the phonon satellite peaks at finite frequency.  This initially results in an increased asymmetry of the spectral function, since only the lowest phonon satellite peaks are populated, resulting 
in the observed increase in the  thermopower with increasing electron-phonon coupling. Eventually, however, for $\lambda/\omega_0\gtrsim 3$, the higher-lying multiphonon
peaks become populated, resulting in a broad distribution of the spectral function centered outside the Fermi window and the thermopower decreases again.
We see that at some intermediate coupling ($\lambda/\omega_0\approx2.0$)  the two trends in the spectral function described above compensate each other and we 
achieve the maximum Seebeck coefficient at this strength of the electron-phonon coupling. For the parameters used in
Fig.~\ref{fig:GSK}(b), we  estimate a maximum Seebeck coefficient of $S_{\rm max}=S(T\approx 7\Gamma_{\rm eff})\approx 2 k_{\rm B}/e=172 \mu V/K$ close to the optimal value of $207 \mu V/K$ found by Sofo and Mahan for
a bulk thermoelectric with a $\delta$ function quasiparticle density of states\cite{Mahan95}.
\footnote{Note, however, that  in Ref.~\onlinecite{Mahan95}, the value $S=207\mu V/K$ was obtained by 
optimizing the dimensionless figure of merit,  including also the lattice contribution to the thermal conductance. They show that the corresponding optimal value for the position of the quasi-particle peak (gate voltage) is 
$\pm 2.4 k_B T$, close to what we found, $\tilde{\varepsilon}_0/T_{\rm max}=-25\Gamma_{\rm eff}/(10\Gamma_{\rm eff})=-2.5$.}

\paragraph{Thermal conductance}
The electronic contribution to the thermal conductance exhibits two peaks, as shown in Fig.~\ref{fig:GSK}(c).
The first peak occurs at temperatures comparable to the renormalized tunneling rate $\Gamma_{\rm eff}$ and reflects the possibility of heat transport via resonant tunneling.
For sufficiently strong couplings ($\lambda/\omega_0>2$), this low-energy peak becomes irrelevant, a consequence
of the decrease in the height and width of the low-energy quasiparticle peak upon increasing the coupling strength [see inset to Fig.~\ref{fig:spectral_function}(a) in Appendix~\ref{subsec:spectral_function}].
The second peak occurs at a temperature related to the phonon frequency $\omega_0$:  it reflects the onset of inelastic scattering processes, which become relevant when the energy of the electrons is sufficient to create or annihilate one or several phonons.
The temperature of this peak position for coupling strengths $\lambda/\omega_0=3,2,1$, and $0.5$ is estimated roughly to be $4.8\omega_0,\; 2.2\omega_0,\;0.75\omega_0$, and $0.45\omega_0$, respectively. For the smallest coupling shown ($\lambda/\omega_0=0.2$), it merges with the lowest peak at $T\approx 0.12\omega_0 \approx 0.6\Gamma_{\rm eff}$. Note also that the position of the second peak in $\kappa_e(T)$ correlates with but is not identical to that in $G(T)$.

\subsubsection{Varying the gate voltage}  
\label{subsubsec:GSK-gate-voltage-dep}
Having seen that in the antiadiabatic regime ($\omega_0=5\Gamma$), and for $\tilde{\varepsilon}_0=-\Gamma$, the thermopower achieves its maximum value at a rather strong electron-phonon coupling $\lambda/\omega_0=2$, 
we want to now keep this optimal coupling strength (and $\omega_0=5\Gamma$) fixed, and investigate further the effect of the gate voltage on the temperature dependence of the various transport properties [Figs.~\ref{fig:GSK}(d)-\ref{fig:GSK}(f)].
\paragraph{Electrical conductance}
As we increase the gate voltage, starting from $|\tilde{\varepsilon}_0/\Gamma|\ll 1$, with $|\tilde{\varepsilon}_{0}|<\Gamma_{\rm eff}$, the low-temperature  electrical conductance is only moderately suppressed 
while the high-temperature conductance remains largely unaffected [see Fig.~\ref{fig:GSK}(d)]. Further increasing
the gate voltage such that $|\tilde{\varepsilon}_{0}|>\Gamma_{\rm eff}$,  leads to  an activated behavior of the conductance, with a maximum at a low temperature which approximately scales with the gate voltage.  Further increase of the gate voltage suppresses the
low-temperature conductance and the maximum further. The second peak in $G(T)$ at higher temperatures, which results from phonon excitations, is independent of the gate voltage.
\paragraph{Seebeck coefficient}
For the Seebeck coefficient, the effect of  increasing the gate voltage away from the particle-hole symmetric point is to first enhance $S(T)$ but for sufficiently large gate voltages $|\tilde{\varepsilon}_{0}|\gg \Gamma_{\rm eff}$ 
there is a decrease in $S(T)$ [see Fig.~\ref{fig:GSK}(e)]. The position of the maximum approximately correlates with $\tilde{\varepsilon}_0$ [compare with the position of the lowest peak in $G(T)$ in Fig.~\ref{fig:GSK}(d)].
These trends in $S(T)$ for varying gate voltage can be qualitatively understood as resulting from a compromise between an increase in the asymmetry of the spectral function and a decrease in the magnitude of the spectral function as we move $\tilde{\varepsilon}_0$ further 
away from the Fermi level resulting in a maximum thermopower for the value  $\tilde{\varepsilon}_0=-\Gamma$ [see Fig.~\ref{fig:GSK}(e) and the spectral function in Fig.~\ref{fig:spectral_function}(d) of Appendix~\ref{subsec:spectral_function}]. 
\paragraph{Thermal conductance}
For the thermal conductance, shown in Fig.~\ref{fig:GSK}~(f), we find similar trends in the gate voltage dependence as in the electrical conductance: a high-temperature peak at a temperature related to $\omega_0$, which is independent of the gate voltage, 
and a much smaller low-temperature peak. This low-temperature peak lies at  $T\approx \Gamma_{\rm eff}$ for $|\tilde{\varepsilon}_0/\Gamma_{\rm eff}|\ll 1$. With increasing gate voltage
$|\tilde{\varepsilon}_0/\Gamma_{\rm eff}|\gg 1$ it shifts to higher
temperatures (correlating with the gate voltage), becomes suppressed, and eventually merges with the high-temperature peak [see inset to Fig.~\ref{fig:GSK}(f)]. 

\subsubsection{Varying the phonon frequency}  
\label{subsubsec:GSK-frequency-dep}
Finally, in Figs.~\ref{fig:GSK}(g)-\ref{fig:GSK}(i), we investigate the effect of the phonon frequency $\omega_0/\Gamma$ on the transport properties, keeping now the optimal coupling strength ($\lambda/\omega_0=2$) and the optimal gate voltage ($\tilde{\varepsilon}_0/\Gamma=-1$) found above. A largely monotonic trend in the transport properties is seen at essentially all temperatures as we go from the adiabatic ($\omega_0\ll\Gamma$) to the antiadiabatic limit ($\omega_0\gg\Gamma$).
We note here that while $\omega_0=\Gamma$ is usually taken as the crossover scale from the adiabatic to antiadiabatic behavior, recent studies\cite{Eidelstein13,Jovchev13}  show that $\omega_0=\Gamma_{\rm eff}$ is a more appropriate definition.
For strong coupling $\lambda/\omega_0\gg 1$, this extends the antiadiabatic regime to significantly lower phonon frequencies. For the results presented below, and those in Sec.~\ref{subsubsec:PZL-frequency-dep}, this means that the 
actual crossover scale between the adiabatic and the (extended) antiadiabatic regime corresponds to $\omega_{0}=0.4\Gamma$ [when $\Gamma_{\rm eff}(\lambda/\omega_0=2)=0.4\Gamma=\omega_0$] and not $\omega_0=\Gamma$.
\paragraph{Electrical conductance}
As we increase the phonon frequency, the low-temperature enhancement of the electrical conductance through the resonant level is suppressed since the resonant tunneling amplitude
$\Gamma_{\rm eff}$ is reduced with increasing $\omega_0$ [see Fig.~\ref{fig:GSK}(g)]. At higher temperatures, a phonon-assisted peak develops in the conductance for large $\omega_0/\Gamma$.
\paragraph{Seebeck coefficient}
Figure~\ref{fig:GSK}(h) shows the monotonic enhancement of the Seebeck coefficient on increasing
$\omega_0/\Gamma$ in the (extended) antiadiabatic regime $\omega_0\geq 0.4\Gamma$ and a small monotonic suppression in the adiabatic regime. The monotonically increasing Seebeck coefficient can 
be qualitatively understood from the behavior of the spectral function with increasing phonon frequency, see  Fig.~\ref{fig:spectral_function}(c) of Appendix~\ref{subsec:spectral_function}.
With increasing $\omega_0/\Gamma$, the lowest energy quasiparticle peak in the spectral function sharpens, becoming more delta-function-like, while remaining asymmetric and located at $\nu\approx \tilde{\varepsilon}_0$
[see Fig.~\ref{fig:spectral_function}(c) in Appendix~\ref{subsec:spectral_function}]. This sharp resonance leads to the monotonic enhancement of the Seebeck coefficient with increasing $\omega_0\gg \Gamma$. Since the quasiparticle
peak in the spectral function occurs at the gate voltage, the temperature of the maximum in the thermopower also correlates with gate voltage and is almost independent of $\omega_0$ [see inset to Fig.~\ref{fig:GSK}(h)].
\paragraph{Thermal conductance}
From Fig.~\ref{fig:GSK}(i), we confirm once more that the high-temperature maximum in the heat transport at a temperature related to $\omega_0$ is 
due to the inelastic phonon-assisted tunneling. On the other hand, the low-temperature heat transport for $T\approx\Gamma_{\rm eff}$ is 
strongly suppressed with increasing phonon frequency [see inset to Fig.~\ref{fig:GSK}(i)]. This very small low-temperature thermal conductance will play a role later when we discuss the figure of merit. 

\subsubsection{Comparison with FRG}
\label{subsubsec:frg-comp-antiadiabatic}
\begin{figure}[t]
 \centering
 \includegraphics[width=1.0\linewidth]{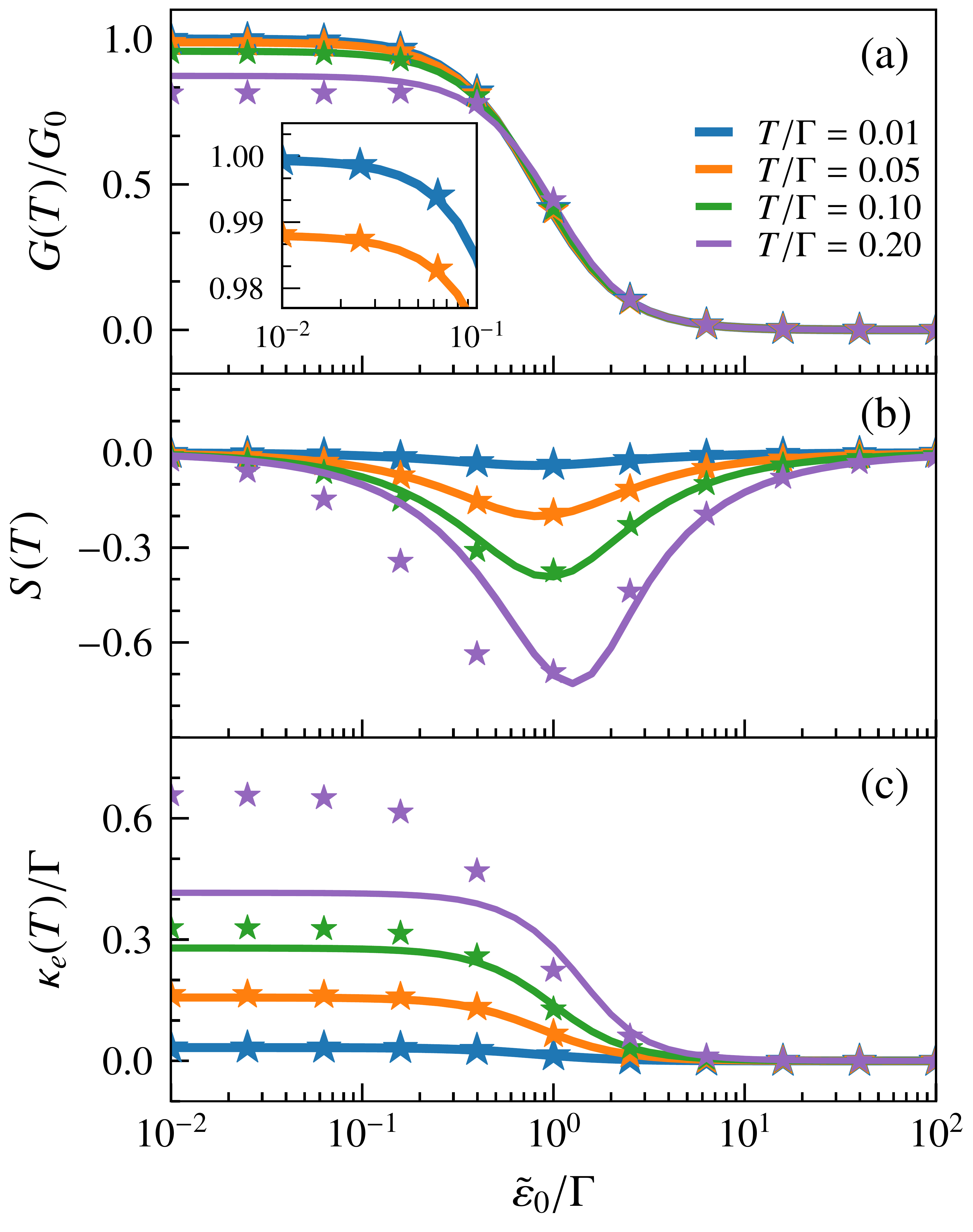}
 \caption{(a) The normalized electrical conductance $G/G_0$, (b) the Seebeck coefficient $S$ (in units of $k_{\rm B}/e$), and, (c), the normalized electronic contribution to the thermal conductance $\kappa_e/\Gamma$ (in units of $k_{\rm B}/h$) vs the dimensionless gate voltage $\tilde{\varepsilon}_0/\Gamma$ for various temperatures
 in the antiadiabatic limit ($\omega_0/\Gamma=20$) for $\lambda/\omega_0=0.5$.
 The solid lines represent the NRG data and the stars are calculated with FRG (using the Sommerfeld expansion).}
 \label{fig:SF}
\end{figure}
\begin{figure*}[t]
 \centering
 \includegraphics[width=1.0\linewidth]{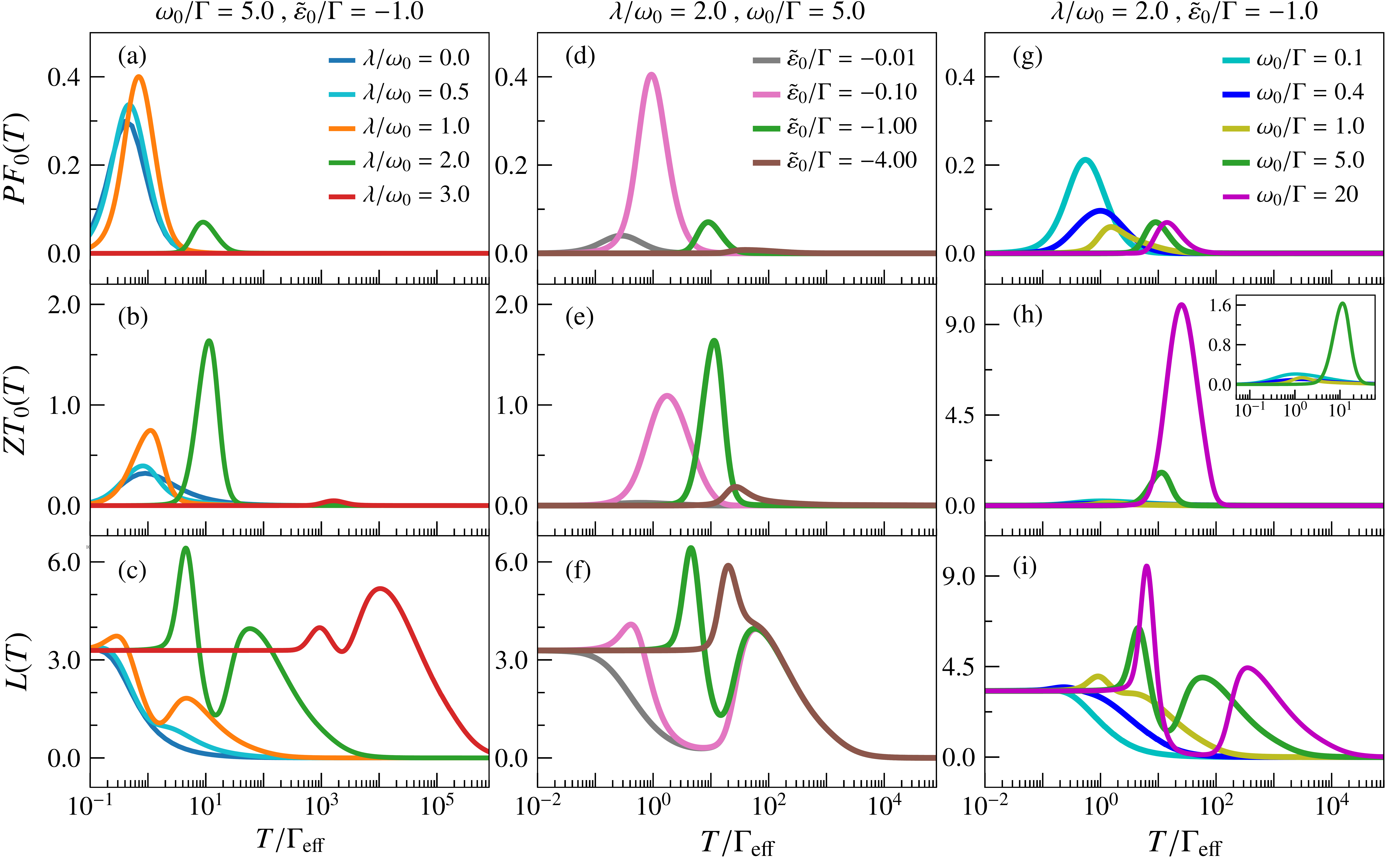}
 \caption{The power factor $PF_0$ (in units of $k_{\rm B}^2/h$), the Lorenz number $L$ (in units of $k_{\rm B}^2/e^2$) and the dimensionless figure of merit $ZT_0$ vs the reduced temperature $T/\Gamma_{\rm eff}$: (a)-(c) for different electron-phonon coupling 
 for a given phonon frequency $\omega_0/\Gamma=5$ and gate voltage $\tilde{\varepsilon}_0=-\Gamma$, (d)-(f) at different gate voltages for a given coupling 
 $\lambda/\omega_0=2.0$ and a fixed frequency $\omega_0/\Gamma=5$, and,  (g)-(i) from the adiabatic to the antiadiabatic limit for a $\lambda/\omega_0=2.0$ and $\tilde{\varepsilon}_0=-\Gamma$.}
 \label{fig:PZL}
\end{figure*}
We briefly return to Figs.~\ref{fig:GSK}(a)-\ref{fig:GSK}(c) and comment on the comparisons between the NRG (lines) and FRG (stars) results shown there. The approach used for calculating these FRG results is based on
Eq.~(\ref{eq:CFmoments}), which uses a continued fraction expansion for the Fermi function appearing in the transport integrals. While this approach works well at all temperatures for the lowest moment, 
and thus for the electrical conductance,  and for couplings up to order $\lambda/\omega_0\approx 1$ [see Fig.~\ref{fig:GSK}(a)], calculating the higher moments within this approach is more problematic. 
The reason is that the first and second moments involve the derivative of the Green's function [see Eq.~(\ref{eq:molpropagator})]. This means that the interpolation scheme used to calculate the transport properties within the FRG 
becomes increasingly sensitive with increasing temperature when the spacing between Matsubara frequencies becomes larger. 
We have tried the P\'ade approximation for the interpolation which turns out to be rather unstable. 
Hence, for the thermopower and thermal conductance, a better approach for the low-temperature regime, $T\ll \Gamma_{\rm eff}$, is to calculate the transport integrals via a Sommerfeld expansion to order $T^2$ 
(see Appendix~\ref{subsec:frg-finite-T}).  Within this approach, the results for the gate voltage dependence of the transport properties at several low temperatures, shown in Fig.~\ref{fig:SF},  agree very well with those calculated from the NRG.  
These low-temperature comparisons also provide an independent check on the NRG calculations. Note, that the deviations at higher temperatures, e.g., for $T/\Gamma=0.2$, which with  $\Gamma_{\rm eff}=0.81\Gamma$ corresponds 
to $T/\Gamma_{\rm eff} \approx 0.25$, are expected since at such temperatures the neglected higher-order terms in the Sommerfeld expansion will
start contributing significantly.

\subsection{Power factor, figure of merit and Lorenz number}
\label{subsec:fom}

To study the interplay between heat and charge transport in more detail, we discuss the temperature dependence of some of the derived thermoelectric quantities such as the power factor $PF_0$,
the dimensionless figure of merit $ZT_0$, and the Lorenz number $L$, defined as
\begin{align}
PF_0(T)=S^2(T)G(T),\\
ZT_0(T)=\frac{G(T)S^2(T)T}{\kappa_e(T)},\\
L(T)=\frac{\kappa_e(T)}{TG(T)}. 
\label{eq:PZL}
\end{align}

As in Sec.~\ref{subsec:transport properties}, we shall discuss the temperature dependence of these quantities for varying
electron-phonon coupling $\lambda$ (at fixed gate voltage $\tilde{\varepsilon}_0$ and phonon frequency $\omega_0$),
gate voltage $\tilde{\varepsilon}_0$ (at fixed coupling $\lambda$ and frequency $\omega_0$) and phonon frequency $\omega_0$ (at fixed coupling $\lambda$ and gate voltage $\tilde{\varepsilon}_0$).  Results for these
three cases are shown in Figs.~\ref{fig:PZL}(a)-\ref{fig:PZL}(c), Figs.~\ref{fig:PZL}(d)-\ref{fig:PZL}(f) and Figs.~\ref{fig:PZL}(g)-\ref{fig:PZL}(i) are be discussed in Secs.\ref{subsubsec:PZL-coupling-dep}-\ref{subsubsec:PZL-frequency-dep}.

\subsubsection{Varying the electron-phonon coupling}
\label{subsubsec:PZL-coupling-dep} 
Figures \ref{fig:PZL}(a)-\ref{fig:PZL}(c) show the temperature dependence of $PF_0$, $ZT_0$, and $L$ for different electron-phonon couplings. 
The power factor as well as the figure of merit exhibit a maximum at a temperature that correlates with $\max\{\Gamma_{\rm eff},|\tilde{\varepsilon}_0|\}$, 
and this maximum is more significant for some rather strong electron-phonon coupling $\lambda/\omega_0=2$
[see Figs.~\ref{fig:PZL}(a)-\ref{fig:PZL}(b)]. \footnote{Note that we used reduced (and not absolute) temperature $T/\Gamma_{\rm eff}$ in Figs.~\ref{fig:PZL}(a) and \ref{fig:PZL}(b), so the position of
the gate-voltage-related peak shifts to higher $T/\Gamma_{\rm eff}$ with increasing  $\lambda$ due to the decrease of $\Gamma_{\rm eff}$ with increasing $\lambda$.}
This maximum is a manifestation of the resonant tunneling and is suppressed as we approach the strong-coupling regime, since
the gate voltage becomes larger than the effective tunneling rate.
The Lorenz number at low temperatures ($T\ll\max\{\Gamma_{\rm eff},|\tilde{\varepsilon}_0|\}$) takes the universal value $L_0=\pi^2k^2_B/3e^2$, reflecting  the Wiedemann-Franz law. The latter states that the ratio of the thermal conductance to the electrical conductance is linear in temperature with proportionality constant $L_0$. In the noninteracting case ($\lambda/\omega_0=0$), as we increase the temperature, the Lorenz number decreases monotonically.
However, in the presence of the phonon-assisted tunneling, the Lorenz number exhibits one low-temperature and one high-temperature maximum. The latter occurs at a temperature related, but not equal, to the phonon frequency $\omega_0$.
The position of the maximum in the figure of merit coincides with the minima in the Lorenz number, indicating temperatures for which the charge transport dominates over heat transport and thus causing enhanced thermoelectric efficiency. This follows from $ZT_0=S^2/L$, i.e., a strong violation of the Wiedemann-Franz law indicated by $L(T)\ll L_0$, together with an enhanced thermopower $S$, favoring an enhanced thermoelectric efficiency.  

\subsubsection{Varying the gate voltage}
\label{subsubsec:PZL-gate-voltage-dep}  
In Figs.~\ref{fig:PZL}(d)-\ref{fig:PZL}(f), we characterize the effect of the gate voltage on the above quantities.
If we apply gate voltages well below or well above $\Gamma_{\rm eff}$, the enhancement of the figure of merit (and the power factor) becomes less substantial [see Figs.~\ref{fig:PZL}(d)-\ref{fig:PZL}(e)].
It is interesting to note that for gate voltages comparable to the effective tunneling rate, see Table~\ref{tab:gamma_eff}, 
the temperatures at which the minimum Lorenz number is realized extends to a rather broad region, as is shown in Fig.~\ref{fig:PZL}(f) for the case $\tilde{\varepsilon}_0=-0.01\Gamma$.

\subsubsection{Varying the phonon frequency}
\label{subsubsec:PZL-frequency-dep}  
Finally, Figs.~\ref{fig:PZL}(g)-\ref{fig:PZL}(i) show the dependence of $PF_0$, $ZT_0$, and $L$ on $\omega_0/\Gamma$. As we approach the antiadiabatic limit, the effective tunneling rate decreases and hence at a finite gate voltage, the resonant tunneling 
is suppressed, resulting in a decrease of the electrical conductance.
The monotonic enhancement of the Seebeck coefficient [cf Fig.~\ref{fig:GSK}(h)] is not sufficient to compensate for the suppression of the electrical conductance [Fig.~\ref{fig:GSK}(g)], and hence the power factor decreases as we increase the phonon frequency [Fig.~\ref{fig:PZL}(g)].    
The figure of merit, on the other hand, increases monotonically with the vibrational frequency once $\omega_0$ exceeds
$\Gamma_{\rm eff}$, i.e., in the extended antiadiabatic limit
[see Fig.~\ref{fig:PZL}(h) and the inset, and, for more details, Fig.~\ref{fig:ZT0scan} of Appendix~\ref{subsec:additional-results}].
The temperature interval for which the enhancement of the figure of merit is realized (and/or the plateaulike region for the minimum Lorenz number) extends as we go to the antiadiabatic limit [Figs.~\ref{fig:PZL}(h)-\ref{fig:PZL}(i)].
In short, in the antiadiabatic limit, when the vibrations are much faster than the tunneling processes, for temperatures $\Gamma_{\rm eff}<T<\omega_0$, the figure of merit is significantly
enhanced and the Lorenz number is substantially suppressed [Fig.~\ref{fig:PZL}(i)]. 
For $\omega_0=5\Gamma$, the maximum value of $ZT_0$ is of order 1 at $T\approx 0.6\Gamma_{\rm eff}$.
The figure of merit continues to grow to even higher values in the extreme antiadiabatic limit $\omega_0\gg \Gamma$ eventually saturating in the limit $\omega_0/\Gamma\to\infty$ [see Fig.~\ref{fig:ZT0scan} of Appendix~\ref{subsec:additional-results}].
However, at such high phonon frequencies, additional vibrational modes or anharmonic effects, neglected in our model, would play a role and invalidate the present description.
In addition, it should be noted that we have neglected lattice phonons of the electronic leads in the spinless Anderson-Holstein model
and hence the computed dimensionless figure of merit $ZT_{0}$ is just an upper bound to the true figure of merit $ZT=S^2GT/(\kappa_e+\kappa_l)$ where $\kappa_l$ is the contribution to the thermal conductance from phonons in the leads and has been neglected here. \cite{Andergassen11} Therefore, in comparing with actual experimental data, the trends that we find may be relevant, but not the exact values for the dimensionless figure of merit. 

Finally, we note also that while the power factor, useful in cooling a hot source,\cite{Mahan98} is large at weak couplings [Fig.~\ref{fig:PZL}(a)],
low finite gate voltages [Fig.~\ref{fig:PZL}(d)], and in the adiabatic regime [Fig.~\ref{fig:PZL}(g)],
the figure of merit, useful in harvesting waste heat, is largest at moderately strong couplings ($\lambda/\omega_0=2$),
finite gate voltages ($\tilde{\varepsilon}_0=-\Gamma$), and in the strongly antiadiabatic regime ($\omega_0\gg\Gamma$).

\section{Summary and outlook}
\label{sec:summary}
\begin{figure*}[t]
 \centering 
 \includegraphics[width=1.0\linewidth]{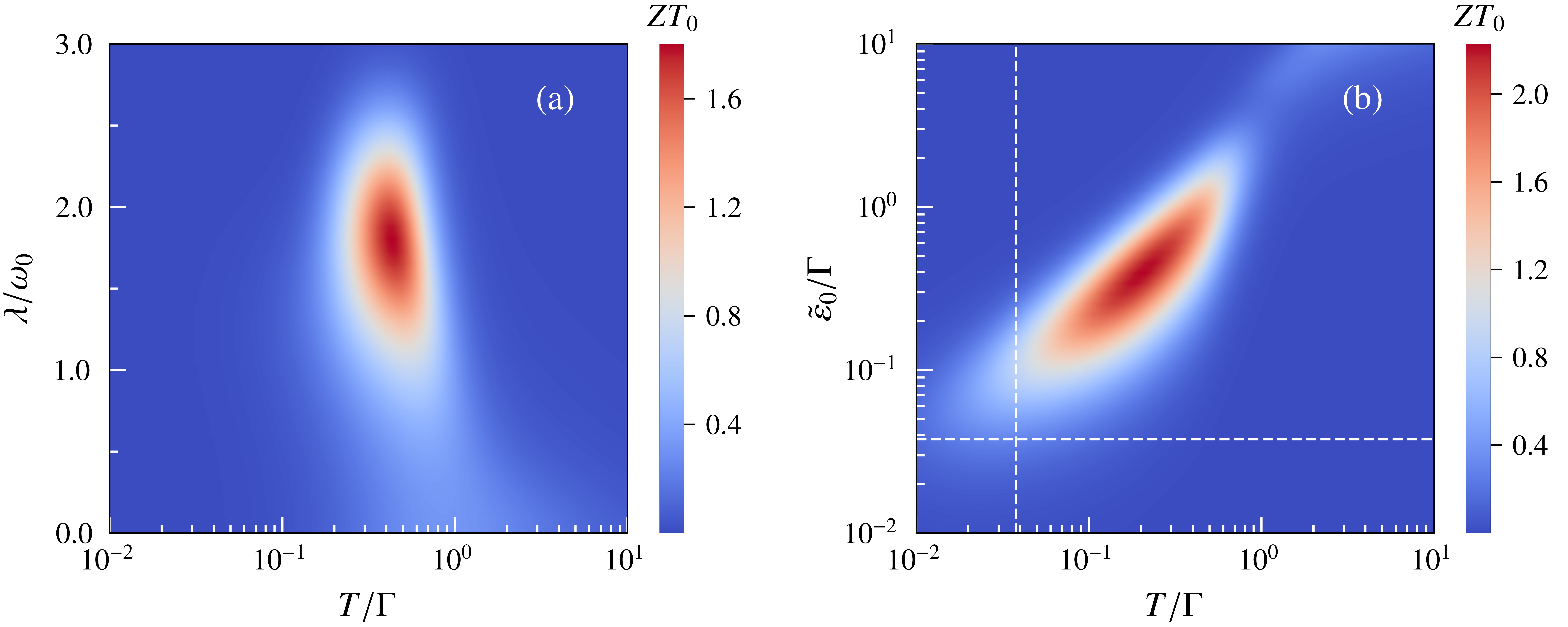}
 \caption{(a) $ZT_0$ vs the coupling strength and temperature for a fixed gate voltage $\tilde{\varepsilon}_0=-\Gamma$ and phonon frequency $\omega_0/\Gamma=5.0$.
(b) $ZT_0$ vs the gate voltage and temperature for a fixed coupling strength $\lambda/\omega_0=2.0$ and phonon frequency $\omega_0/\Gamma=5.0$. Vertical and horizontal dashed lines in (b) indicate $T=\Gamma_{\rm eff}$ and $|\tilde{\varepsilon}_0|=\Gamma_{\rm eff}$, respectively.}
 \label{fig:ZTcolor}
\end{figure*}
We studied the effect of the vibrational degrees of freedom on the linear thermoelectric transport through a molecular quantum dot described by the spinless Anderson Holstein model by using the NRG method.
As an independent check, we compared them to corresponding finite temperature transport calculations within the FRG approach for weak to intermediate couplings at different gate voltages
and in the antiadiabatic regime. We found that the emergent low-energy scale $\Gamma_{\rm eff}$ and the phonon-assisted tunneling play important roles in understanding the thermoelectric transport at finite temperatures.
We quantified the trends in the transport properties in the adiabatic and antiadiabatic regimes. 
In the antiadiabatic regime, we showed that strong electron-phonon coupling, induces at finite gate voltages, an asymmetry in the spectral function. 
This results in an enhancement of the Seebeck coefficient, and thereby yields another route to enhanced thermoelectric efficiency in molecular quantum dots with vibrational degrees of freedom, 
which is akin to the Mahan-Sofo mechanism for bulk thermoelectric materials. 

Figures~\ref{fig:ZTcolor}(a)-\ref{fig:ZTcolor}(b) summarize the parameter regimes for which an enhanced dimensionless figure of merit is realized.
In Fig.~\ref{fig:ZTcolor}(a), for a fixed gate voltage $\tilde{\varepsilon}_0/\Gamma=-1.0$ and in the antiadiabatic regime ($\omega_0/\Gamma=5.0$), we 
see an enhanced $ZT_0\gtrsim 1$ for temperatures $0.3\Gamma \lesssim T\lesssim 0.7\Gamma$ and couplings $1.25\lesssim \lambda/\omega_0 \lesssim 2.25$. 
Using typical values for $\Gamma=10~{\rm meV}$ and $\omega_0=5\Gamma = 50~{\rm meV}$,\cite{Galperin07} we find a temperature range of $30~{\rm K}\lesssim T \lesssim 70~{\rm K}$ for enhanced thermoelectric efficiency.
In Fig.~\ref{fig:ZTcolor}(b), we keep the coupling strength fixed to $\lambda/\omega_0=2.0$ (in the optimal range) and look at the variation
of $ZT_0$ as a function of the gate voltage and temperature. We see an enhanced figure of merit $ZT_0\gtrsim 1$ for temperatures $0.04\Gamma\lesssim T \lesssim 0.6\Gamma$, 
and for gate voltages $0.08\Gamma\lesssim|\tilde{\varepsilon}_0|\lesssim 1.6\Gamma$, or upon using $\Gamma_{\rm eff}\approx 0.04\Gamma$ from Table~\ref{tab:gamma_eff},
for temperatures  $\Gamma_{\rm eff}\lesssim T \lesssim 15\Gamma_{\rm eff}$, and for gate voltages $2\Gamma_{\rm eff}\lesssim\tilde{\varepsilon}_0\lesssim 40\Gamma_{\rm eff}$.
These correspond to  temperatures and gate voltages in the ranges $4~{\rm K} < T <60~{\rm K}$ and $0.8 ~{\rm meV}<|\tilde{\varepsilon}_0|< 16~{\rm meV}$, respectively, upon
using $\Gamma=10~{\rm meV}$ and $\omega_0=5\Gamma = 50~{\rm meV}$.
We expect that similar enhancements in $ZT_0$ can be found within the spinful Anderson-Holstein model 
in the regime of a weak local Coulomb repulsion on the dot. For larger Coulomb repulsion, we expect that 
spin Kondo physics will suppress the observed enhancement in the thermoelectric efficiency.

In the future, we plan to include the repulsive electron-electron interaction between the molecular dot and the leads in a further step to make the model more realistic.
As found earlier,\cite{Kennes13} models with such short-range Coulomb interactions exhibit in their nonequilibrium (steady-state) thermoelectric transport some nontrivial and intriguing features leading to an enhancement of their thermoelectric efficiency.
In this light, we plan also to go beyond linear response theory and investigate nonequilibrium thermoelectric transport through a molecular quantum dot, including vibrational and short-range Coulomb terms
within an FRG approach on the Keldysh contour.
The advantage of the latter, beyond being applicable to both nonlinear and linear transport, is that it can be carried out directly on the real energy axis,
thereby avoiding problems with the analytic continuation of numerical data.

\begin{acknowledgments} 
This work was supported by the Deutsche Forschungsgemeinschaft via RTG 1995. 
We acknowledge supercomputer support by the John von Neumann Institute for Computing (J\"ulich). 
One of us (T.A.C.) acknowledges the hospitality of the Aspen Center for Physics, supported by the National Science Foundation through Grant No. PHY-1607611, where part of this work was carried out.
\end{acknowledgments} 
\appendix
\section{FRG at finite temperatures}
\label{subsec:frg-finite-T}
To set up FRG in Matsubara space at finite temperatures, we use the cutoff function as has been introduced in Ref.~\onlinecite{Enss05}:
\begin{equation}
\Theta_T\big(|\nu_n|-\Lambda\big)=
\begin{cases}
0 &|\nu_n|-\Lambda\le-\pi T\\
\frac{1}{2}+\frac{|\nu_n|-\Lambda}{2\pi T}&\Big||\nu_n|-\Lambda\Big| < \pi T\\
1 & |\nu_n|-\Lambda\ge \pi T\\
\end{cases}
\label{eq:T_dep_cuoff}
\end{equation}
Following, the standard procedure within first-order truncated FRG,\cite{Metzner12,Khedri17} we obtain coupled differential equations for the real and the imaginary part of the self-energy
[$\Sigma^{\Lambda}(i\nu_m)=\epsilon^{\Lambda}(i\nu_m)+i\gamma^{\Lambda}(i\nu_m)$]:
\begin{align}
\partial_{\Lambda}\epsilon^{\Lambda}(i\nu_m)=\frac{1}{\beta}\mbox{Re}\big\{S^{\Lambda}(i\nu_{\tilde{n}})\big\}
\sum_{s=\pm} [U(i\nu_m-si\nu_{\tilde{n}})-U(0)],
\label{eq:flowrefiniteT}
\end{align}
\begin{align}
\partial_{\Lambda}\gamma^{\Lambda}(i\nu_m)&=\frac{1}{\beta}\mbox{Im}\big\{S^{\Lambda}(i\nu_{\tilde{n}})\big\}
\sum_{s=\pm} sU(i\nu_m-si\nu_{\tilde{n}}),
\label{eq:flowimfiniteT}
\end{align}
where $\tilde{n}$ is the integer for which the corresponding Matsubara frequency $\nu_{\tilde{n}}\in(\Lambda-\pi T,\Lambda+\pi T)$ at given temperature $T$ and a scale factor $\Lambda$.
The effective phonon-mediated electron-electron interaction is
\begin{align}
U(i\nu_n)=-\frac{2\omega_0\lambda^2}{\nu_n^2+\omega_0^2},
\end{align}
and the single-scale propagator $S^{\Lambda}(i\nu_{\tilde{n}})$ reads
\begin{align}
S^{\Lambda}(i\nu_{\tilde{n}})=\frac{i\nu_{\tilde{n}}+i\Gamma\operatorname{sgn}(\nu_{\tilde{n}})}{\big[i\nu_{\tilde{n}}
+i\Gamma\operatorname{sgn}(\nu_{\tilde{n}})-\alpha(T,\Lambda)\Sigma^{\Lambda}(i\nu_{\tilde{n}})\big]^2}\nonumber\\
\times \frac{\Theta\Big(\pi T-\big||\nu_{\tilde{n}}|-\Lambda\big|\Big)}{2\pi T},
\end{align}
with $\alpha(T,\Lambda)=\frac{1}{2}+\frac{|\nu_{\tilde{n}}|-\Lambda}{2\pi T} \in (0,1)$.
The initial conditions at $\Lambda\to\infty$ are
\begin{equation}
\epsilon^{\Lambda}(i\nu_n)=\epsilon_0-E_{\rm p}=\tilde{\varepsilon}_0\hspace{0.2cm},\hspace{0.2cm} \gamma^{\Lambda}(i\nu_n)=0 \hspace{0.5cm} \forall \nu_n. 
\end{equation}
At the particle-hole symmetric point $\epsilon_0=E_{\rm p}$, the real part will not flow, reflecting that the particle-hole symmetry is preserved at any temperature for all scales $\Lambda$.
We used standard adaptive routines to numerically solve the flow equations.

At low temperatures $T\ll\Gamma$, we can calculate the transport integrals Eq.~(\ref{eq:transportintegrals}) using the Sommerfeld expansion
\begin{align}
I_n=\pi\Gamma\bigg[F_n+\frac{\pi^2}{3}\frac{F^{\prime\prime}_n}{2!\beta^2}\bigg], 
\end{align}
where
\begin{align}
F_n=\delta_{n,0}\frac{-1}{\pi}\text{Im}\big\{G(i\nu_1)\big\},
\end{align}
\begin{align}
F^{\prime\prime}_n=&\frac{2}{\pi}\delta_{n,1}\text{Im}\{G_{\rm mol}^2(i\nu_1)\}\bigg(1-\text{Im}\bigg\{\frac{d\Sigma(i\nu_n)}{d\nu_n}\bigg|_{\nu_1}\bigg\}\bigg)\nonumber\\
&+\delta_{n,0}\frac{-1}{\pi}\Bigg[2\text{Im}\big\{G_{\rm mol}^3(i\nu_1)\big\}\bigg(1-\text{Im}\bigg\{\frac{d\Sigma(i\nu_n)}{d\nu_n}\bigg|_{\nu_1}\bigg\}\bigg)^2\nonumber\\
&+\text{Im}\big\{G_{\rm mol}^2(i\nu_1)\big\}\bigg(\text{Re}\bigg\{\frac{d^2\Sigma(i\nu_n)}{d\nu_n^2}\bigg|_{\nu_1}\bigg\}\bigg)\Bigg]\nonumber\\ 
&+\delta_{n,2}\frac{-2}{\pi}\text{Im}\{G_{\rm mol}(i\nu_1)\},
\end{align}
with $\nu_1=\pi/\beta$.
\begin{figure}[t]
 \centering 
 \includegraphics[width=1.0\linewidth]{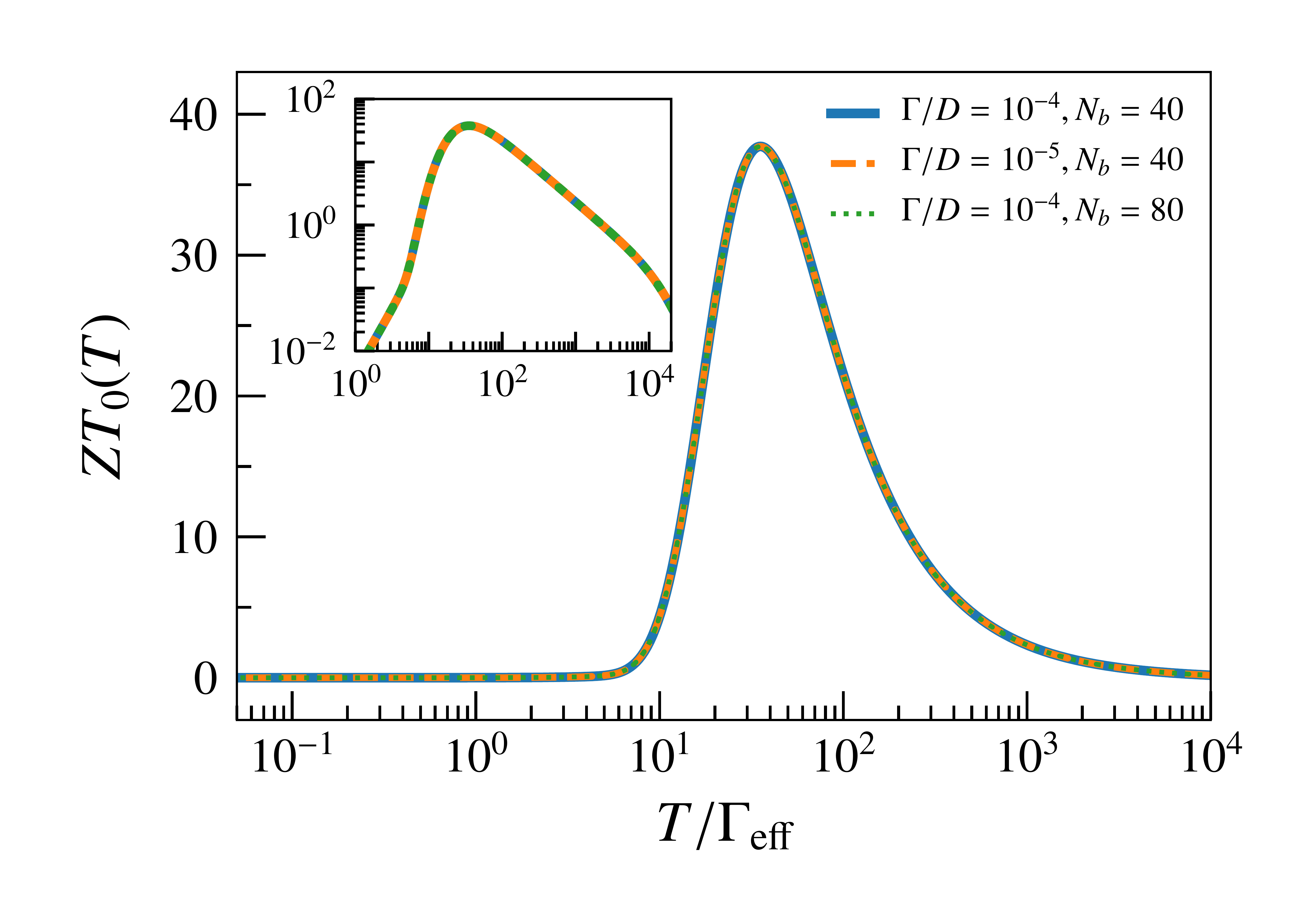}
 \caption{The dimensionless figure of merit $ZT_0$ vs reduced temperature $T/\Gamma_{\rm eff}$ for different number of bosons and different $\Gamma/D$ deep in the antiadiabatic regime 
 $\omega_0/\Gamma=10^3$ for $\lambda/\omega_0=2.0$,$\tilde{\varepsilon}_0=-\Gamma$, and illustrating the convergence with respect to the number of bosons kept.}
 \label{fig:con}
\end{figure}
\begin{figure}[t]
\centering 
\includegraphics[width=1.0\linewidth]{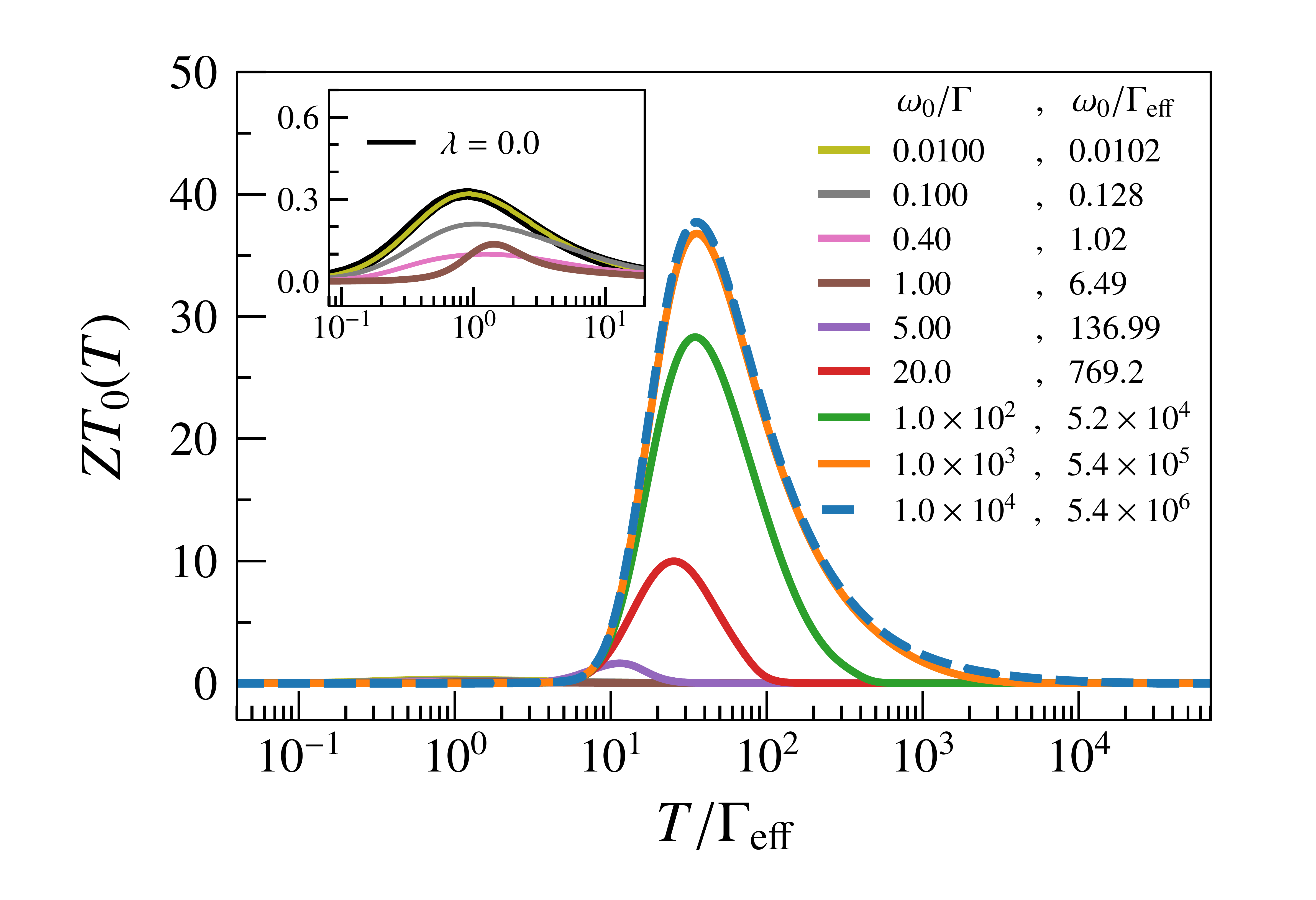}
\caption{The dimensionless figure of merit $ZT_0$ vs the reduced temperature $T/\Gamma_{\rm eff}$ for different $\omega_0/\Gamma$ at $\tilde{\varepsilon}_0=-\Gamma$
and $\lambda/\omega_0=2$. The values of $\omega_0/\Gamma_{\rm eff}$ are also listed. The main panel shows the monotonically increasing behavior of the peak value of $ZT_0$ 
with increasing $\omega_0/\Gamma$ in the extended antiadiabatic regime ($\omega_{0}/\Gamma_{\rm eff}\gtrsim 1$), 
while the inset shows the opposite behavior for $\omega_0/\Gamma_{\rm eff}\lesssim 1$.
The $\lambda=0$ curve in the inset lies on top of the $\omega_0/\Gamma=0.01$ curve, illustrating that the noninteracting
and adiabatic limits are almost identical.}
\label{fig:ZT0scan}
\end{figure}
\begin{figure*}[t]
\includegraphics[width=0.96\textwidth]{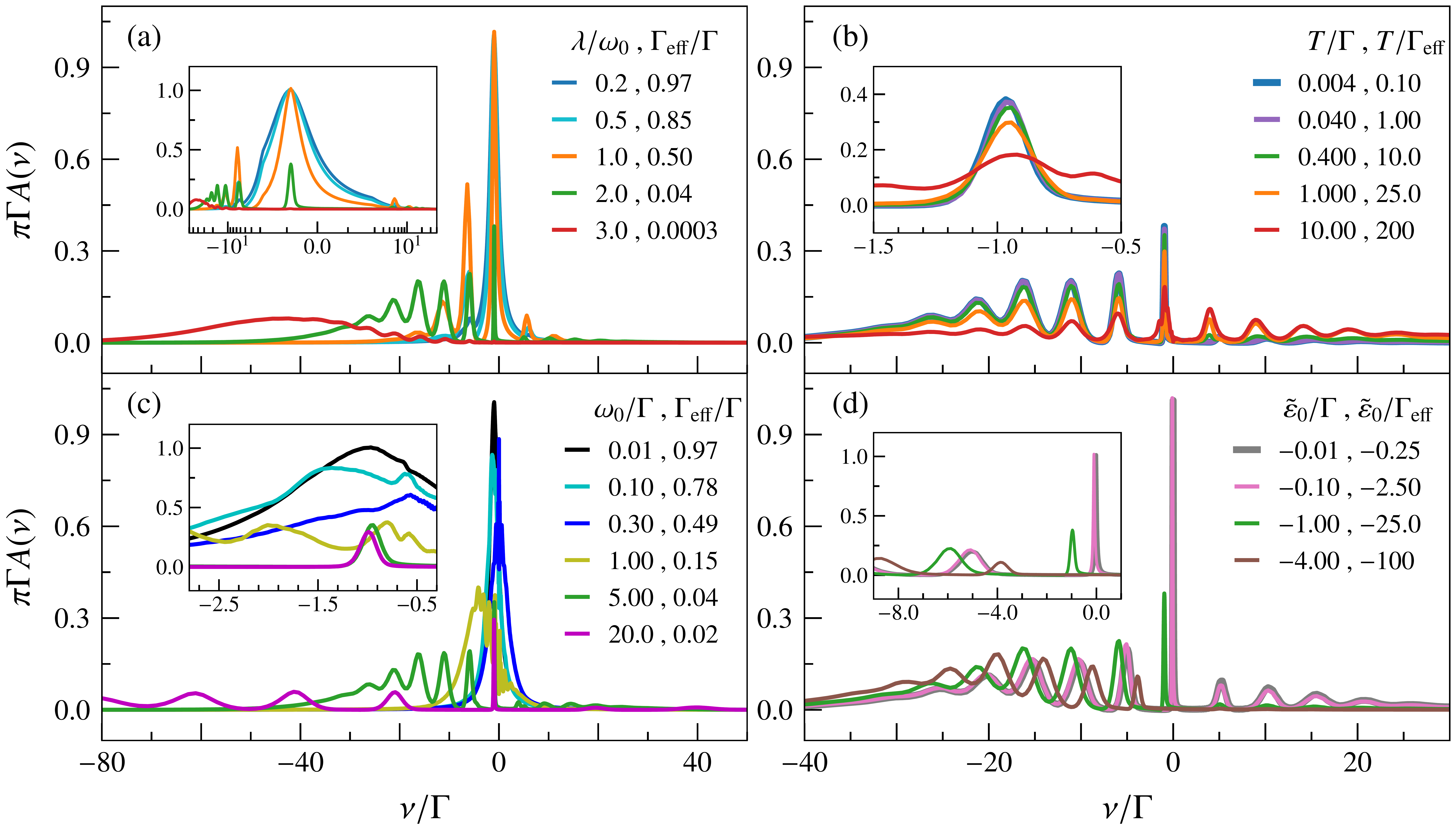}
\caption{The normalized spectral function $\pi\Gamma A(\nu)$ vs frequency $\nu$.
(a) The coupling strength dependence of the $T=0$ spectral function for $\tilde{\varepsilon}_0=-\Gamma$ 
and $\omega_0=5\Gamma$.
(b) The temperature dependence of the spectral function for $\tilde{\varepsilon}_0=-\Gamma$, $\omega_0=5\Gamma$ and $\lambda/\omega_0=2$.
(c) The (phonon) frequency dependence of the spectral function for $\tilde{\varepsilon}_0=-\Gamma$ 
, $\lambda/\omega_0=2.0$ and $T/\Gamma=0.4$. (d) The gate voltage dependence of the $T=0$ spectral function for $\omega_0=5\Gamma$ and $\lambda/\omega_0=2$. 
We used NRG parameters $\Lambda=4$ and $N_z=20$. The spectral sum rule $\int_{-\infty}^{+\infty}A(\nu)d\nu=1$ is satisfied numerically to within a few percent for all parameters and temperatures shown.}
\label{fig:spectral_function}
\end{figure*} 
\section{NRG parameters}
\label{subsec:NRG-parameters} 
We always check the convergence of the presented results with respect to the number $N_b$ of bosons kept.
Figure~\ref{fig:con} shows an example for the extreme antiadiabatic regime, where the phonon excitations are significant 
to capture the physics for some rather strong electron-phonon coupling $\lambda/\omega_0=2.0$. One sees that results for $N_b=40$ are indistinguishable from those for $N_b=80$ at all temperatures. Note also that using a smaller $\Gamma=10^{-5}$ did not require more than $N_b=40$ phonons for converged results.
\section{Trends in $ZT_0$ versus $\omega_0/\Gamma$ for strong coupling}
\label{subsec:additional-results}
For completeness, and in order to further elucidate on the trends previously observed, we show results for the dimensionless figure of merit as a function of $\omega_0/\Gamma$
extending up to very large $\omega_0/\Gamma$ in Fig.~\ref{fig:ZT0scan}. We see that while initially for $\omega_0/\Gamma_{\rm eff} \lesssim 1$, the peak 
value of the dimensionless figure of merit decreases with increasing $\omega_0/\Gamma$ (see inset to Fig.~\ref{fig:ZT0scan}) for $\omega_0/\Gamma_{\rm eff} \gtrsim 1$ 
its peak value exhibits a monotonically increasing behavior (main panel in Fig.~\ref{fig:ZT0scan}). 
This maximum value, eventually saturates to approximately $35$ for $\omega_0/\Gamma \gg 1$. This maximum is located at a temperature $T/\Gamma_{\rm eff}\approx 30$.

\section{Spectral function}
\label{subsec:spectral_function}

The thermopower directly probes the asymmetry of the spectral function about the Fermi level, so the dependence of this asymmetry on parameters such as the electron-phonon coupling strength, 
the temperature, the phonon frequency, and the gate voltage can give some qualitative insight into the observed trends of the Seebeck coefficient. 
Hence, we discuss these dependences in this Appendix.

Figures~\ref{fig:spectral_function}(a)-\ref{fig:spectral_function}(d) show the spectral function $A(\nu)$ vs $\nu$ upon varying  $\lambda/\omega_0$ (for $\omega_0/\Gamma=5$ and $T=0$) 
, $T/\Gamma$ (for $\lambda/\omega_0=2$ and $\omega_0/\Gamma=5$), $\omega_0/\Gamma$ (for $\lambda/\omega_0=2$ and $T/\Gamma=0.4$), and $\tilde{\varepsilon}_0/\Gamma$
(for $\omega_0/\Gamma=5$ and $\lambda/\omega_0=2$).

In Fig.~\ref{fig:spectral_function}(a), one sees how the asymmetry in the spectral function develops with increasing $\lambda/\omega_0$ with additional phonon satellite peaks appearing at $\lambda/\omega_0\gtrsim 1$.
The resulting asymmetry in the spectral function, with sharp peaks at $\nu\approx\tilde{\varepsilon}_0-\omega_0, \tilde{\varepsilon}_0-2\omega_0,\dots$,
qualitatively explains the monotonically increasing thermopower with increasing $\lambda/\omega_0\lesssim 2$.
Eventually, however, for $\lambda/\omega_0 \gtrsim 2$ these satellite peaks acquire less weight and the first moment of the spectral function starts to decrease resulting in a reduction of the thermopower at very strong coupling.
This qualitatively explains the trends seen in Fig.~\ref{fig:GSK}(b).

The spectral function shown in
Fig.~\ref{fig:spectral_function}(a) is obtained by broadening the discrete spectra with logarithmic Gaussians using broadenings proportional to the excitation energies of the delta peaks.\cite{Bulla08}
At strong coupling the phonon excitations merge into a broad peak centered at a large negative frequency for the broadening parameters used here.
One can resolve the individual satellite peaks under this broad feature by further reducing the broadening, see Ref.~\onlinecite{Jovchev13}.
However, in the actual transport calculations reported in Sec.~\ref{sec:results}, we work directly with the discrete spectra
using Eq.~\ref{eq:NRGmoments}, and the results for the transport properties do not depend on any broadening procedure, which are hence highly accurate for all temperatures.\cite{Oliveira09} 

For the optimal parameters of Fig.~\ref{fig:spectral_function}(b), the Seebeck coefficient shows a maximum for $\Gamma_{\rm eff}\lesssim T \lesssim\Gamma$, where the asymmetric resonant tunneling is realized within the Fermi window.
As we increase the temperature further, the phonon side peaks in the spectral function are increasingly broadened, as shown in Fig.~\ref{fig:spectral_function}(b). In addition,
the asymmetry in the spectral function is reduced. The  latter reflects the fact that thermal excitations involving emission of  phonons ($\nu\approx n\omega_0$) are becoming as relevant as those involving absorption of phonons  ($\nu\approx -n\omega_0$) as the temperature is increased.

Figure~\ref{fig:spectral_function}(c) shows the dependence of the spectral function on phonon frequency for 
optimal coupling $\lambda/\omega_0=2$ and at temperature $T/\Gamma=0.4$ [where the Seebeck coefficient has its maximum value for the case $\omega_0/\Gamma =5$ in Fig.~\ref{fig:GSK}(e)].
For this case of strong coupling, the spectral function retains a large weight in the peak close to the Fermi level upon increasing $\omega_0/\Gamma$ [see also the inset to 
Fig.~\ref{fig:spectral_function}(c)].  This low-energy peak, which can be identified with the quasiparticle peak in the limit $T\to 0$, remains pinned at the gate voltage $\tilde{\varepsilon}_0=-\Gamma$ and sharpens with
increasing phonon frequency. The resulting asymmetry explains the monotonic increase of the low-temperature thermopower with increasing phonon frequency in Fig.~\ref{fig:GSK}(h) for $\omega_{0}\gtrsim \Gamma_{\rm eff}$. In addition,
its pinning at the gate voltage $\nu=\tilde{\varepsilon}_0=-\Gamma$ explains why the maximum in the thermopower occurs at a temperature correlating with the gate voltage and largely independent of $\omega_0$
[inset to Fig.~\ref{fig:GSK}(h)].

Figure~\ref{fig:spectral_function}(d) shows the gate voltage dependence of the $T=0$ spectral function. For $|\tilde{\varepsilon}_0/\Gamma|\ll 1$, the main contribution to the low-temperature thermopower comes from the quasiparticle peak
at low energies, which is seen to be quite symmetrical about the Fermi level, thereby resulting in a small thermopower. Increasing $\tilde{\varepsilon}_0$ shifts this peak away from the Fermi level, increasing the asymmetry of the spectral function 
within the Fermi window and thereby increasing the thermopower. Eventually, for sufficiently large gate voltage, most weight will lie outside the Fermi window and the thermopower will decrease. The optimal thermopower is found for
$\tilde{\varepsilon}_0=-\Gamma$ (for the chosen $\omega_0=5\Gamma$ and $\lambda/\omega_0=2$). 
\begin{figure}[t]
\centering 
\includegraphics[width=1.0\linewidth]{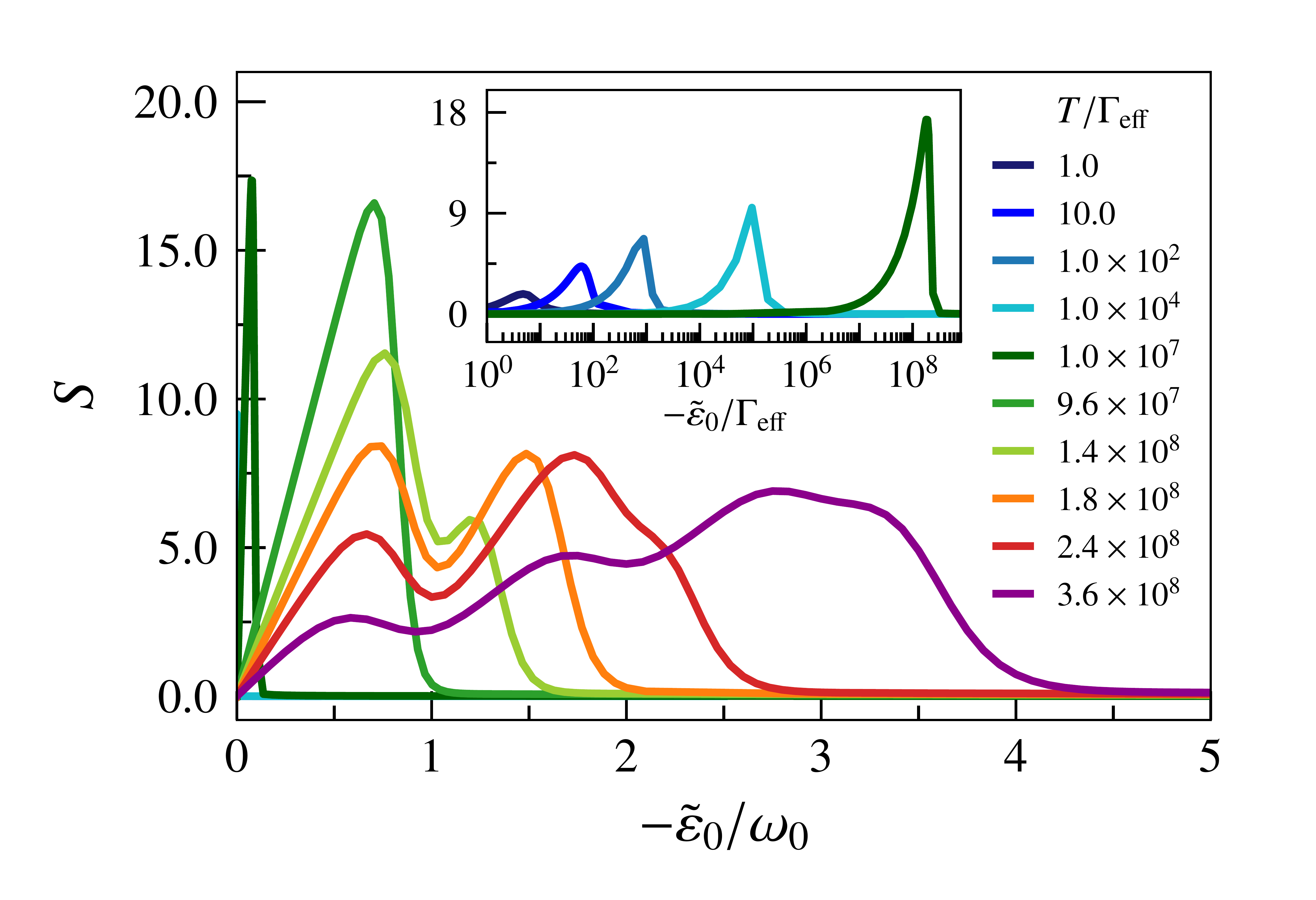}
\caption{Thermopower $S$ vs the normalized gate voltage $-\tilde{\varepsilon}_0/\omega_0$ for various temperatures, in the extreme antiadiabatic limit $\omega_0/\Gamma=4.0\times10^{7}$
and for a fixed electron-phonon coupling $\lambda/\omega_0=2.0$. The inset shows $S$ vs $-\tilde{\varepsilon}_0/\Gamma_{\rm eff}$ for the five lowest temperatures of the main panel.
We choose $\Gamma/D=10^{-10}$ to resolve high-temperature results more precisely. 
The renormalized tunneling rate is $\Gamma_{\rm eff}=1.659\times 10^{-2}\Gamma$.
}
\label{fig:SV_gTscan}
\end{figure}
\section{Evolution of the thermopower versus gate voltage from its high temperature perturbative limit to its low 
temperature strong coupling limit}
\label{subsec:weakGamma}
In this Appendix, we present the gate voltage dependence of the thermopower deep in the antiadiabatic limit $\omega_0/\Gamma\gg 1$,
for different temperatures and show how it evolves as we approach the nonperturbative low-temperature regime $T\lesssim\Gamma_{\rm eff}$ from the high-temperature perturbative one $D\gg T\gg\Gamma >\Gamma_{\rm eff}$.
The latter regime can also be accessed within a rate equation approach, valid for weak lead-molecule couplings, in which only sequential and cotunneling processes are included. \cite{Koch04} In contrast, the NRG, which accounts for all tunneling processes, is able to capture the whole temperature range. 

Starting from low temperatures $T=\Gamma_{\rm eff}$, one sees a single peak in $S(\tilde{\varepsilon}_0)$ at a gate voltage related to the temperature (inset to Fig.~\ref{fig:SV_gTscan}).
This is similar to the low temperature results in the main text, see Fig.~\ref{fig:SF}(b). For sufficiently high temperatures $T\gg\Gamma > \Gamma_{\rm eff}$, and within a small
temperature window  $10^8\lesssim T/\Gamma_{\rm eff}\lesssim 4\times 10^8$ (corresponding to  $0.04 \lesssim T/\omega_{0}\lesssim 0.15$), a number of peaks appear in $S$, approximately separated by integer multiples of $\omega_0$,
which can be attributed to signatures of molecular vibrations in the thermopower. 
This result, for the spinless Anderson-Holstein model, is similar to that found for the
spinful Anderson-Holstein model (including a local Coulomb repulsion) within the rate equation approach of Ref.~\onlinecite{Koch04},
and demonstrates the ability of the NRG to access the high-temperature limit in addition to  accessing low temperatures. On further increasing the temperature,
the peaks in $S$ become shoulders and eventually merge to form a smooth hump whose height decreases with further increase of temperature (not shown).
\bibliography{ref2}
\end{document}